\documentclass[reprint,
aps,pra,    
10pt,
superscriptaddress,
amsmath,amssymb,
floatfix
]{revtex4-1} % For final two-column format

\usepackage{amsmath,amssymb}
\usepackage{mathtools}
\usepackage{esvect}
\usepackage[svgnames]{xcolor}
\usepackage[colorlinks=true,
linkcolor=MediumBlue,
urlcolor=MediumBlue,
citecolor=MediumBlue]{hyperref}
\usepackage{anyfontsize}
\usepackage{titlesec}

\renewcommand{\vec}[1]{\ensuremath{\mathbf{#1}}}

\makeatletter
\renewcommand*{\fnum@figure}{{\footnotesize \figurename~\thefigure}}
\makeatother

\titlespacing\section{0pt}{15pt plus 2pt minus 2pt}{6pt plus 2pt minus 2pt}
\titlespacing\subsection{0pt}{15pt plus 2pt minus 2pt}{6pt plus 2pt minus 2pt}

\begin{document}

\title{Three-dimensional force-field microscopy with optically levitated microspheres}

\author{Charles P. Blakemore}
\email{cblakemo@stanford.edu}
\affiliation{Department of Physics, Stanford University, Stanford, California 94305, USA}

\author{Alexander D. Rider}
\affiliation{Department of Physics, Stanford University, Stanford, California 94305, USA}

\author{Sandip Roy}
\affiliation{Department of Physics, Stanford University, Stanford, California 94305, USA}

\author{Qidong Wang}
\affiliation{Institute of Microelectronics of the Chinese Academy of Sciences, Beijing 100029, China}

\author{Akio Kawasaki}
\affiliation{Department of Physics, Stanford University, Stanford, California 94305, USA}
\affiliation{W. W. Hansen Experimental Physics Laboratory, Stanford University, Stanford, California 94305, USA}

\author{Giorgio Gratta}
\affiliation{Department of Physics, Stanford University, Stanford, California 94305, USA}
\affiliation{W. W. Hansen Experimental Physics Laboratory, Stanford University, Stanford, California 94305, USA}

\date{\today}
\begin{abstract}

We report on the use of 4.7${\text -}\mu$m-diameter, optically levitated, charged microspheres to image the three-dimensional force field produced by charge distributions on an Au-coated, microfabricated Si beam in vacuum. An upward-propagating, single-beam optical trap, combined with an interferometric imaging technique, provides optimal access to the microspheres for microscopy.  In this demonstration, the Au-coated surface of the Si beam can be brought as close as ${\sim}10~\mu$m from the center of the microsphere while forces are simultaneously measured along all three orthogonal axes, fully mapping the vector force field over a total volume of ${\sim}10^6$~$\mu$m$^3$.   We report a force sensitivity of $(2.5 \pm 1.0) \times 10^{-17}~{\rm N / \sqrt{Hz}}$, in each of the three degrees of freedom, with a linear response to up to ${\sim}10^{-13}~{\rm N}$.  While we discuss the case of mapping static electric fields using charged microspheres, it is expected that the technique can be extended to other force fields, using microspheres with different properties. \\

\noindent DOI: \href{https://doi.org/10.1103/PhysRevA.99.023816}{10.1103/PhysRevA.99.023816}

\end{abstract}

\maketitle

\section{INTRODUCTION}

\begin{figure*}[t!!!!]
\includegraphics[width=1.9\columnwidth]{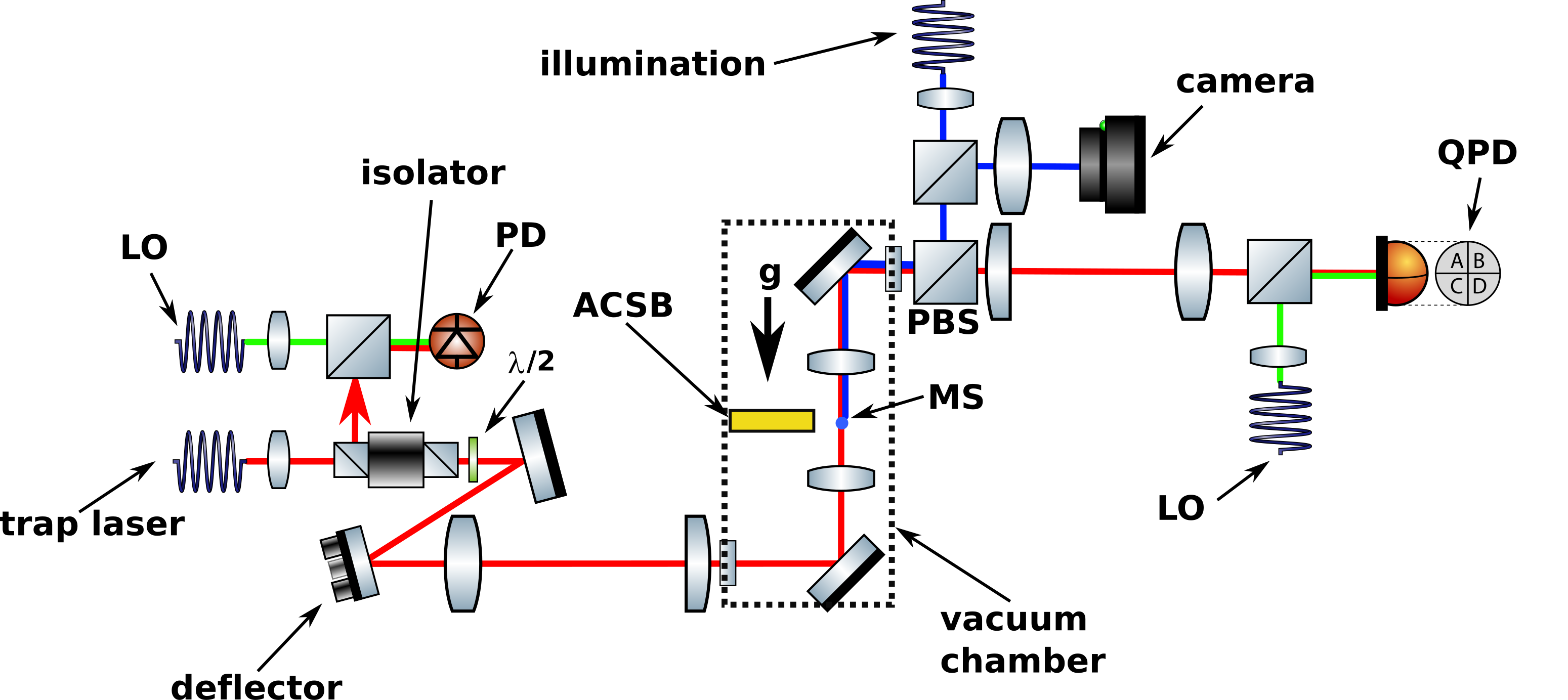}
\caption{ \footnotesize Schematic view of the free-space optical system. The output of the optical fiber carrying the trapping beam is first collimated, then deflected by a high-bandwidth ($f_{\rm -3dB}~{\sim}1~$kHz) piezo-mounted mirror in the Fourier plane of the trap, which produces translations at the trap position. Two aspheric lenses inside the vacuum chamber focus the trapping beam and re-collimate the light transmitted through the MS. The transmitted beam is then interfered, on a quadrant photodiode (QPD), with a local oscillator (LO) beam, whose frequency is shifted by 500~kHz with respect to the trapping beam.  The frequency shift is made prior to free space launch by two fiber-coupled acousto-optic modulators (not pictured), one for the trapping beam and one for both LO beams. Light backscattered by the MS is extracted, combined with another LO beam and used to interferometrically measure the axial position of the MS.  All components are only shown schematically and are not drawn to scale. PD, photodiode; PBS, polarizing beam splitter; and $\lambda/2$, half-wave plate.}
\label{fig:freespace_system}
\end{figure*}

The ability to make measurements at ever smaller length scales has had profound implications for both fundamental science and technology. In particular, atomic force microscopy has enabled the measurement and manipulation  of surfaces at atomic length scales. Traditionally, atomic force microscopes (AFMs) have sensed the interaction of a tip, suspended by a cantilever, with a surface, by measuring the displacements of the cantilever in the direction perpendicular to its surface~\cite{AFM:1986}. The mechanical suspension of the force-sensing element limits electrical, thermal, and mechanical isolation from the outside world.   

Here we present a technique for measuring three-dimensional forces over a three-dimensional volume by levitating a 4.7${\text -}\mu$m-diameter dielectric microsphere (MS) at the focus of a Gaussian laser beam, with a ${\sim}10{\text -}\mu$m minimum distance between the center of the MS and another object.   This results in the full mapping of a vector field anywhere in space and close to mechanical objects. The optical levitation enables three important features: The absence of dissipation associated with the cantilever support allows measurements with substantially lower force noise at room temperature; the electric isolation provided by the optical support makes electrostatic measurements at fixed charge possible; and force vectors measured in three dimensions are characterized by similar spring constants in each of the three trapping degrees of freedom (DOFs). 

Since the pioneering work of Ashkin~\cite{Ashkin:1970,Ashkin:1971,Ashkin:1977}, a number of experiments with optically levitated MSs have been demonstrated, especially in recent years~\cite{Bishop:2004,Chang:2010,Li:2011,Gieseler:2012,Li:2013,Yin:2013,Asenbaum:2013,Moore:2014,Ranjit:2015,Millen:2015,Rider:2016,Jain:2016,Hoang:2016,Ranjit:2016,Mazilu:2016,Fonseca:2016,Vovrosh:2017,Monteiro:2017,Hempston:2017,Rider:2018,Monteiro:2018,Diehl:2018}. Some authors have proposed short-range force detection experiments using their optically levitated MSs~\cite{Geraci:2010,Ether:2015}, but only a few have actually positioned free-standing objects micrometers away from a trapped MS~\cite{Rider:2016,Diehl:2018,Winstone:2018} or other mesoscopic objects. Positioning a Si beam or other attractor close to a trapped MS and measuring its effect on the trapped MS are crucial for any short-range force-sensing application.

As a demonstration of this technique, we measure the electrostatic forces between a charged MS and an Au-coated Si beam (ACSB) structure. A force calibration, obtained by charging the MS with a unity charge and applying an external electric field, allows us to precisely determine the electric field due to a overall bias on the ACSB as well as infer the distribution of ``patch potentials'' on the Au surface in absolute terms.

\section{EXPERIMENTAL SETUP}

The optical trap used here is almost identical to that described in Ref.~\cite{Rider:2018}. The optical system is shown schematically in Fig.~\ref{fig:freespace_system}.  The trap is formed at the focus of an upward-propagating Gaussian beam generated by a 1064-nm laser, whereby the radiation pressure and Earth's gravity create a stable three-dimensional harmonic trap.  Silica MSs of $4.71\pm0.05~\mu$m diameter~\cite{bangs_laboratories} are applied to the bottom side of a glass coverslip, where they adhere by van der Waals forces.   The MSs are loaded into the trap by vibrating the glass slide, placed about 10~mm above the trap, with a piezoelectric transducer.  For this operation, the vacuum chamber is maintained at a pressure of 2~mbar of N$_2$ buffer gas to slow the fall of the MSs and provide sufficient damping of their motion within the trap.   After loading one MS, the glass slide is withdrawn in order to minimize distortions to the trapping beam, which is also used for applying optical feedbacks and measuring forces.  The N$_2$ gas is then gradually pumped out, reaching a final pressure of ${\sim}10^{-6} \, {\rm mbar}$ for the measurements.  

\begin{figure}[b!]
\includegraphics[width=1.0\columnwidth]{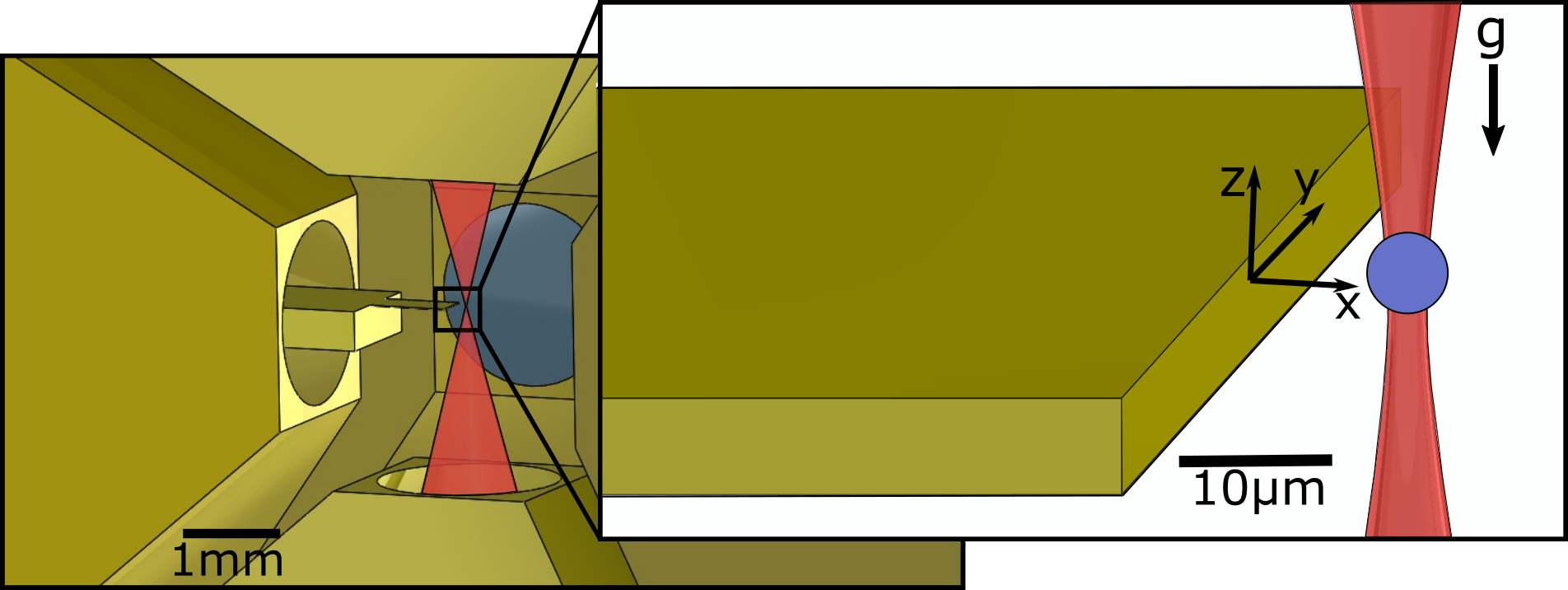}
\caption{ \footnotesize Trap region: A drawing of the trap region is shown in the left panel, illustrating the 2-mm-diameter holes in the Au-coated pyramidal shielding electrodes through which the ACSB and the trapping beam are brought in and extracted.   The trapping light is represented by the red conical feature.   The panel to the right illustrates a detail of the end of the ACSB and trapped MS, along with the coordinate frame used in the data analysis.  A separate arrow shows the direction of Earth's gravity.}
\label{fig:trap}
\end{figure} 

The trap becomes unstable below pressures of ${\sim}0.1~{\rm mbar}$, because of insufficient gas damping. To stabilize the trap under high vacuum conditions, optical feedback forces are applied to the MS. The radial DOFs are stabilized using a piezoelectrically actuated deflection mirror placed in a Fourier plane of the trap, while the axial DOF is stabilized by modulating the power of the laser with an acousto-optic modulator.  The optical readout system used in the feedback for the three DOFs closely follows Ref.~\cite{Rider:2018} with significant improvements made to the electronics. The single-beam configuration of the apparatus is ideal for the application described here, as it allows optimal access to the MS from all directions in the horizontal plane.  The current noise level on the motion of the MSs is dominated by effects~\cite{Rider:2018} other than the residual vacuum, which is limited at $10^{-6}$~mbar by the out-gassing of a number of translation stages. 

The trap region is illustrated in Fig.~\ref{fig:trap}.  To minimize long-range electrostatic interactions, this region is shielded inside a Faraday cage consisting of six hollow, pyramidal electrodes that can be independently biased. The holes in the electrode faces, used to allow for optical access, do not significantly affect the shielding achieved at the trap, based on results from finite-element analysis (FEA). The two electrodes on the vertical axis hold the aspheric lenses, while shielding the trap region from their, potentially charged, dielectric surfaces.  MSs loaded in the trap are generally charged.  Ultraviolet light from a xenon flash lamp, brought in the trap region by an optical fiber, is used to discharge the MSs, while their charge state is monitored by applying an ac voltage to two opposing pyramidal electrodes. As demonstrated in Ref.~\cite{Moore:2014}, this process can be used to verify with extremely high confidence that the MS is overall neutral.  

The force sensitivity of the system is calibrated for each MS with the same process, typically when a charge excess of $1e$ is reached, where $e$ is the elementary charge.  The calibration procedure, described in Refs.~\cite{Rider:2018, Rider:2016, Moore:2014}, can independently measure the force response along the three coordinate axes, by employing, in turn, different electrode pairs.  Small corrections from the fringe field due to the shape of the electrodes are calculated and accounted for using finite-element calculations and the full model of the electrode arrangement. The MS response is frequency dependent, with a natural, radial resonant frequency of ${\sim}400~$Hz, and is measured over the range $1{\text -}600$~Hz. 

Once a force calibration has been obtained, the charge state of the MS can be arbitrarily set for the purpose of electrostatic force sensing.  Continued exposure to UV flashes produces a net positive charge, while flashing the same UV light onto a nearby Au surface produces free electrons, some of which become bound to the MS, generating a net negative charge. Both charging mechanisms occur simultaneously, although the relative rates, and thus the charging direction, can be controlled by bringing the ACSB (an Au surface) close to the MS, or removing it to a distant location.  This flexibility and the demonstrated indefinite stability of charge states afford a large dynamic range in force-sensing applications. For electric field mapping discussed here, the MS is charged to approximately $q\approx-400e$, with an excess of electrons.  This large charge is intended to overwhelm possible multi-polar (mainly dipolar) effects, clearly visible for neutral MSs.

We assume this charging procedure does not produce a large permanent dipole moment via localized charging of the MS. A dipole moment of $d \approx 500e \times 5~\mu$m (aligned for maximal coupling to the field gradient) produces a force that is an order of magnitude smaller than the force on a charge of $q \approx 500e$ in the electric field configuration employed here (discussed below).

\section{MEASUREMENTS AND RESULTS}

The ACSB is designed to have contrast in density (and baryon number) for a future experiment. This is achieved by alternating Au and Si fingers with 25~$\mu$m distance between the centers of contiguous fingers.   A 3${\text -}\mu$m-thick Si fence bridges all Si fingers together, so that the Au fingers are entirely surrounded by Si. A 200-nm Au coating is then applied to all surfaces.   A more complete description of the ACSB is provided elsewhere~\cite{Wang:2017}, while, for the purpose of the work described here, only this external layer of Au is relevant. The ACSB is 10~$\mu$m thick to minimize its interference with the tails of the Gaussian trapping beam. More work on both shaping of the trapping beam and the MS imaging system are likely to make thicker field-generating components possible; nevertheless, constraints of this type are likely to remain the main limitation of this new technique.

The ACSB can be biased independently from other components and is mounted off a three-axis, piezoelectric translational nanopositioning stage (Newport NPXYZ100SGV6).  After trap loading and force calibration, the ACSB is brought close to the trap region through one of the holes in the pyramidal electrodes, as shown in Fig.~\ref{fig:trap}, where the coordinate system employed here is defined.  Sufficient clearance is provided for the full 80${\text -}\mu$m range of the translation stage in the $y$ and $z$ directions. 

Force measurements are carried out in three configurations: to evaluate the noise and linearity of the measurement, to demonstrate the ability of the technique to map the field produced by an external bias on the ACSB, and to map the field produced by patch potentials on the ACSB surface. In all measurements involving the ACSB, the trap is held at a fixed location, while the ACSB is displaced and/or actively driven.

\subsection{Sensor linearity and noise}

\begin{figure}[t!]
\includegraphics[width=1\columnwidth]{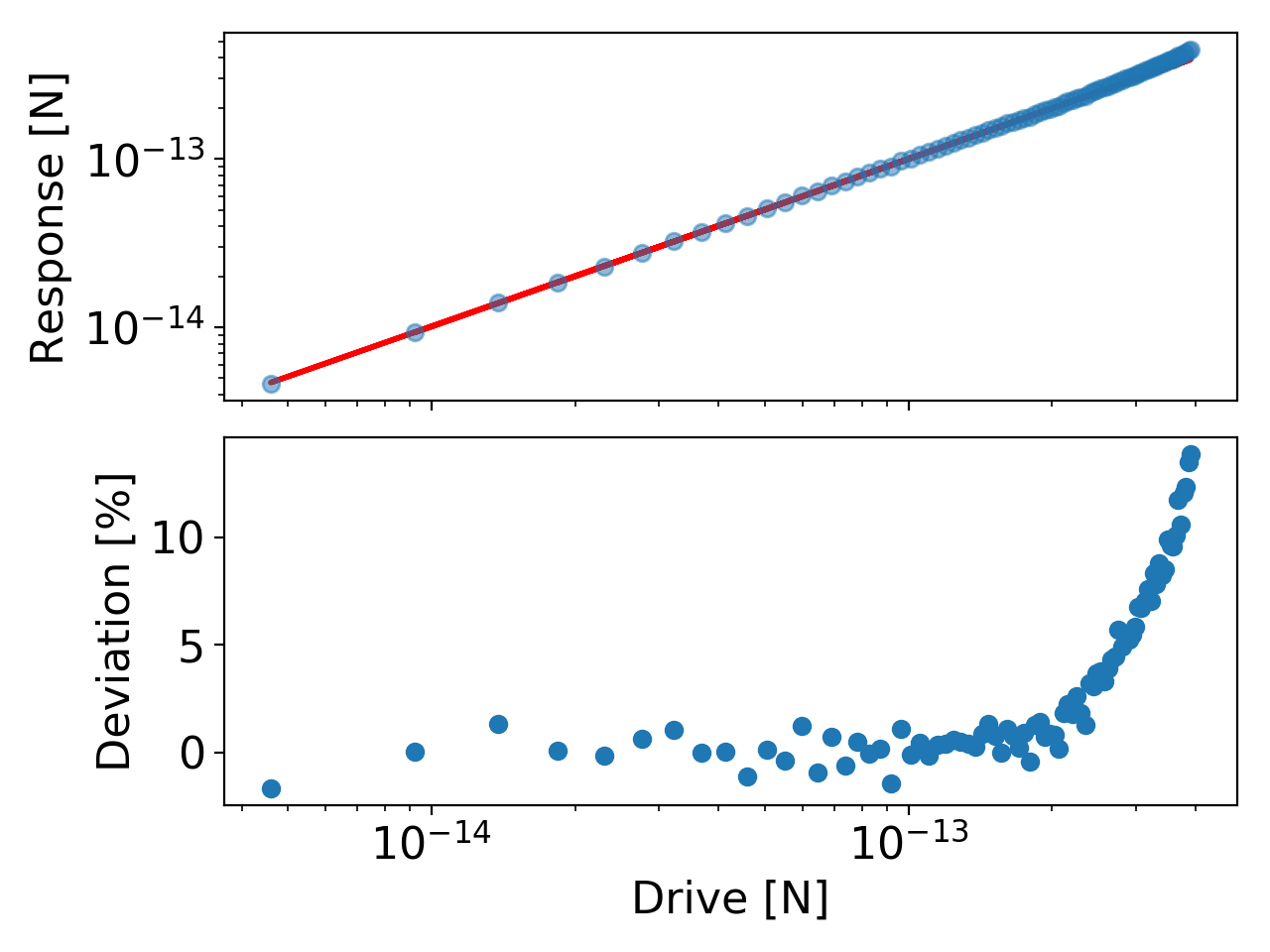}
\caption{ \footnotesize Linearity of the force sensor in the $x$ direction, with residuals from perfectly linear behavior. Data are collected for a 41-Hz drive signal applied with a shielding electrode. The other two DOFs have comparable linearity.}
\label{fig:linearity}
\end{figure}

There are two features that define the performance of a force sensor: noise and linearity.   The linearity of the force measurement is obtained, along the three DOFs, by the same process used for the force calibration, scanning a 41-Hz ac drive signal with the shielding electrodes, from $(0\pm0.1)$ to $500$~fN. The linearity has frequency dependence, which was measured as part of the calibration, as discussed in Sec.~II.  The result, for forces along the $x$ axis, is shown in Fig.~\ref{fig:linearity}, along with the residuals to a perfectly linear behavior.  At driving forces over 300~fN, a ${\sim}$10\% nonlinearity is observed.

\begin{figure}[t!!!]
\includegraphics[width=1.0\columnwidth]{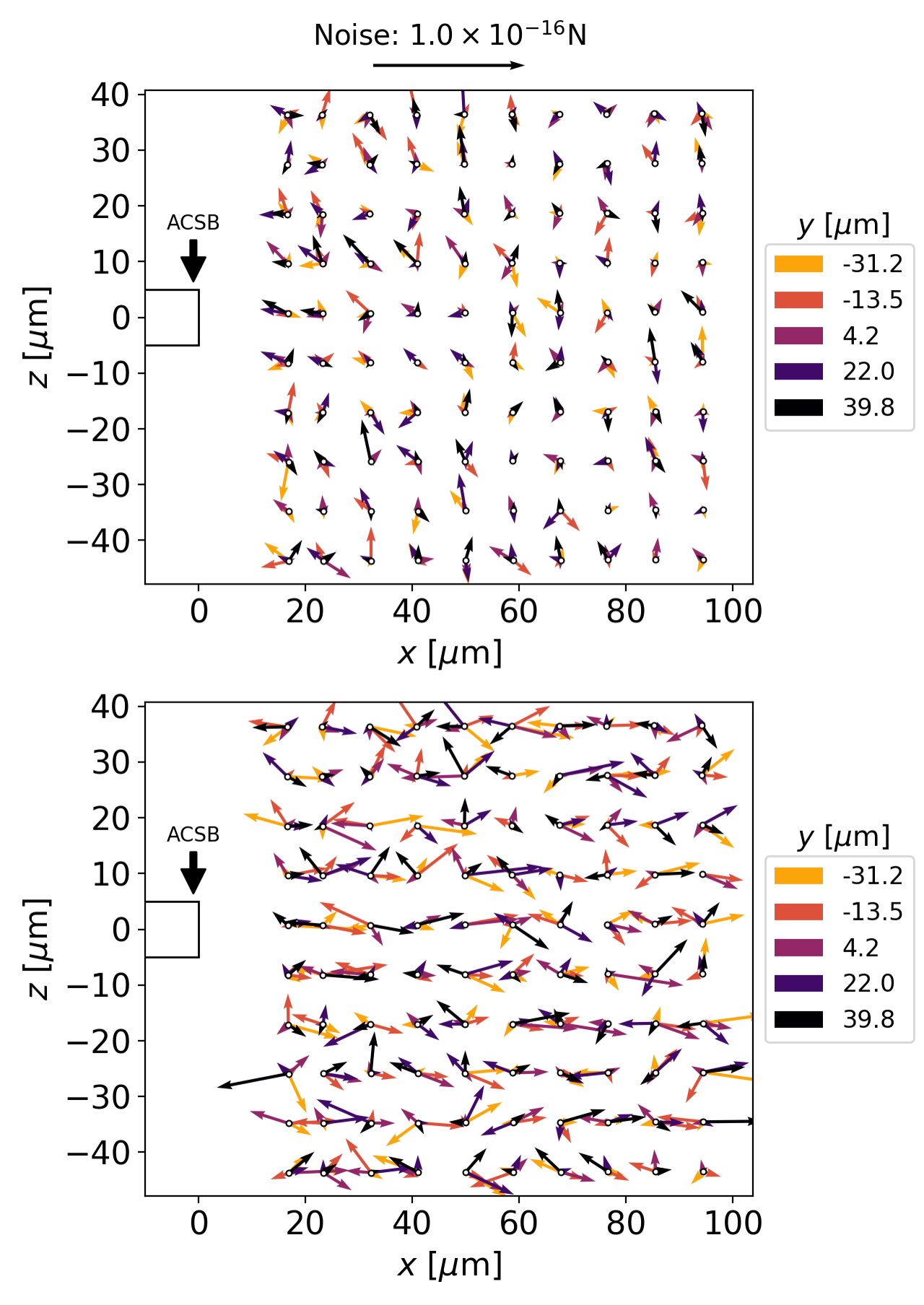}
    \caption{ \footnotesize Force noise micrographs with a charged MS, represented with respect to a stationary ACSB at various relative positions. Practically, changes in the relative position are produced by displacing the ACSB, yet for ease of representation, the figure shows the ACSB as stationary. The force vectors represent the MS response at a frequency far from a single-frequency AC voltage drive applied to the ACSB during these measurements. For the data in the top (bottom) panel, the horizontal component of the force vector represents $F_{x}$, the force in the $x$ direction ($F_{y}$, the force in the $y$ direction), while the $z$ component of the vector is always the force in the $z$ direction. In both cases, force vectors measured at five $y$ positions (with a different color indicating each $y$ position shown in the legend) in a regularly spaced grid ranging from $-40$ to $40~\mu$m are plotted at each $(x,z)$ location.}
    
\label{fig:noise_micrographs}
\end{figure}

\begin{figure}[t!!!!]
\includegraphics[width=1.0\columnwidth]{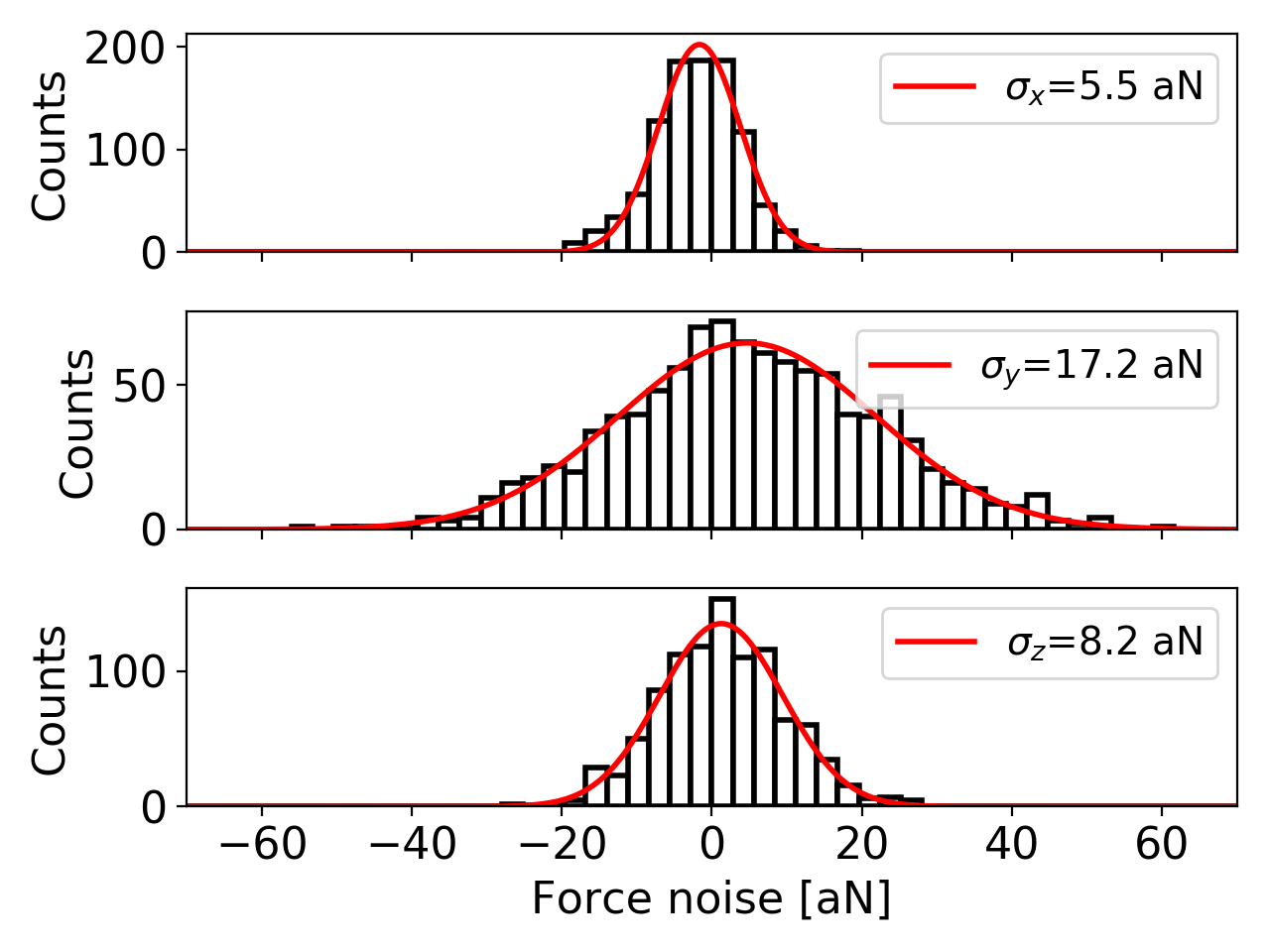}
\caption{ \footnotesize From top to bottom: histograms of the measured force noise in the $x$, $y$, and $z$ directions at different grid locations. Also shown are the fits to normal distributions. The low noise and high linearity of the apparatus enable measurements with a dynamic range of over four orders of magnitude.}
\label{fig:noise_hist}
\end{figure}

The noise is characterized by collecting data with the ACSB placed over a three-dimensional grid of positions in front of a trapped and charged MS. The six trapping electrodes are nominally grounded and the ACSB is driven with a single-frequency ac voltage at $41~$Hz. We then examine the response of the MS at a frequency far from this applied tone. We observe similar noise conditions when the ACSB is nominally grounded.   The resulting force micrographs are shown in Fig.~\ref{fig:noise_micrographs} for 10-s integrations at each position and with a single MS. The rms noise forces over the whole $80\times 80\times 80~\mu$m$^3$ measurement volume are 5.5, 17.2, and 8.2~aN, for the $x$, $y$, and $z$ directions, respectively. Histograms of the noise over the whole measurement volume are plotted in Fig.~\ref{fig:noise_hist}. The anisotropy in the measured noise may be the result of residual astigmatism of the trapping beam, inconsistencies in $x$ and $y$ feedback, or, possibly, effects related to the finite geometry of the ACSB~\cite{Turchette:2000,Low:2011}. This force noise is comparable to a cryogenic Si cantilever~\cite{Chiaverini:2003,Churnside:2014}, but is obtained at room temperature and on all three DOFs simultaneously.  This noise performance, along with the linearity up to ${\sim}10^{-13}~$N, enables force measurements spanning over four orders of magnitude in amplitude.

\subsection{Electric field from an overall bias voltage}

The overall electric field from a bias voltage applied to the ACSB is measured.  This procedure is used here to validate the technique, but also to register the relative position of the ACSB to the trap and its orientation in space, with a fit to a model produced by FEA. In a new version of the trap, currently under construction, high-quality metrology will be possible through auxiliary optics, something only available in a rudimentary fashion in the current system.

\begin{figure}[t!!!!]
\includegraphics[width=1.0\columnwidth]{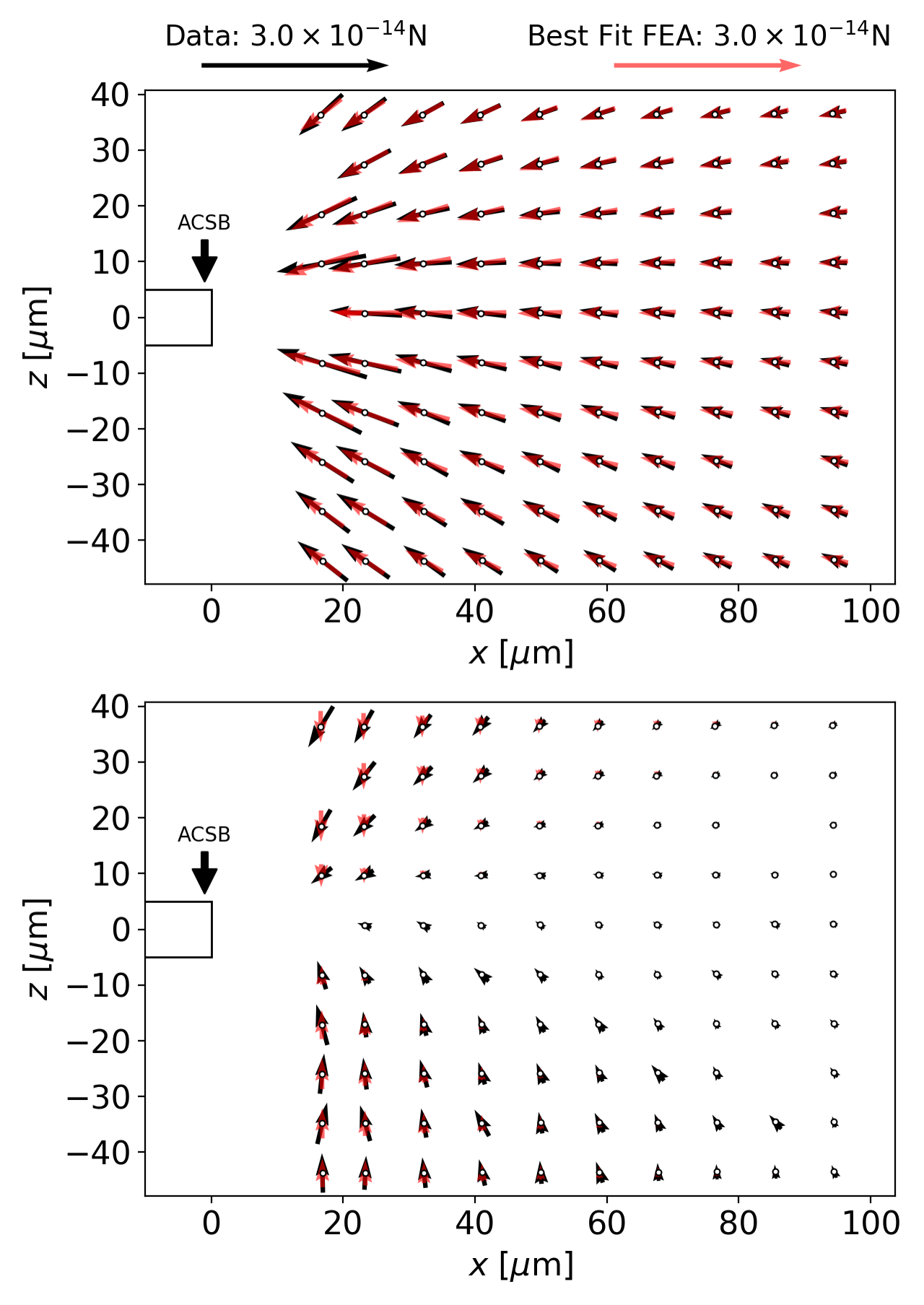}
\caption{ \footnotesize Vector field plots of $(F_x,F_z)$, top, and $(F_y,F_z)$, bottom, in an $xz$ plane of relative positions. As before, changes in the relative position are produced by displacing the ACSB, yet for ease of representation, the figure shows the ACSB as stationary. The black arrows represent the measured force, while the red arrows represent the best fit from the FEA of the $\vec{E}$ field produced by an ACSB with the same overall bias voltage as in the experiment. A few grid points are missing, due to data corruption introduced by the data acquisition instrumentation.}
\label{fig:efield_fit}
\end{figure}

The three-dimensional electric field is mapped over a $10 \times 10 \times 10$ grid of points, spanning the full $80~\mu$m of closed-loop travel in the translation stage, along each of its three orthogonal axes. The relative displacement between points in the three-dimensional scan is known with an uncertainty of ${\sim}10~$nm, set by the accuracy of the translation stage used. To perform the measurement at each point on this grid, the MS is driven for $10~$s by an ac voltage on the ACSB at $41~$Hz,  with a $100{\text -}$mV peak-peak amplitude.  The force at each grid point is represented by a three-dimensional vector. A slice of this vector field along an $xz$ plane at $y=0$ (centered along the $y$ axis) is shown in Fig.~\ref{fig:efield_fit}, together with the results of the FEA. Shown are both $F_x$ and $F_z$ (top) and $F_y$ and $F_z$ (bottom) in the same $xz$ plane.

The data are fit to the FEA model by constructing a least-squared cost function from the difference between them, normalized by the error in the data, and summed over all 1000 grid locations. In the minimization, the charge of the microsphere, $q$ is allowed to float (which is equivalent to applying an overall scaling of the $\vec{E}$ field produced by ACSB), as well as three translations of the coordinates reported by the stage, and six independent rotation angles. Three of these angles floated in the fit represent angular misalignment between the axes of the translation stage and the axes of the trap, which are used to construct a rotation matrix applied to the measurement grid points. The remaining three angles account for a possible angular misalignment of the ACSB itself to the axes of the translation stage, and are used to construct a rotation matrix applied to the measured vector field. We exclude constant offsets in the measured force, as would arise from contact potentials, since we apply an ac electric field and measure the amplitude and phase of the MS's response.

With this procedure, we find $q=-459e$ (consistent with the estimate made during the charging process), and a closest separation of the ACSB face of $15~\mu$m from the center of the MS, centered in the $z$-axis with an uncertainty $\pm2.5~\mu$m.  The coordinate axes of the ACSB are found to be tilted relative to the coordinate axes of the translation stage by no more than $\pm5^{\circ}$ for all three rotation angles, while the coordinate axes of the translation stage are tilted relative to the physical axes of the trap by no more than $\pm2^{\circ}$ for all three rotation angles. 

Residual deviations from the best-fit FEA, particularly apparent at short separations, are likely the result of non-uniformity of the voltage on the ACSB, as our FEA assumes a perfect conductor with ideal geometry.  Small dielectric particles (e.g., silica dust) on the surface of the ACSB, and/or metallic grains in the Au coating, may contribute to the small discrepancies observed at shorter distances~\cite{Rossi:1992,Robertson:2006}. Permanent dipole moments in the MS, estimated from Ref.~\cite{Rider:2016}, may produce forces more than an order of magnitude smaller than the residual force from the fit described above. Similarly, induced dipoles, estimated from the MS index of refraction \cite{bangs_laboratories} and the contact potential on the ACSB assumed to be $<100~{\rm mV}$, may produce a force more than two orders of magnitude smaller than the residuals. 

The closest approach of the ACSB, as described above, is $15~\mu$m for the dataset shown in Fig.~\ref{fig:efield_fit}. Force sensing at smaller distances, down to ${\sim}8~\mu$m, is possible and has been achieved with this technique, but with an increase of the system noise for which a satisfactory explanation has not been found.

\subsection{Patch potential measurements}

\begin{figure}[t!]
\includegraphics[width=1.\columnwidth]{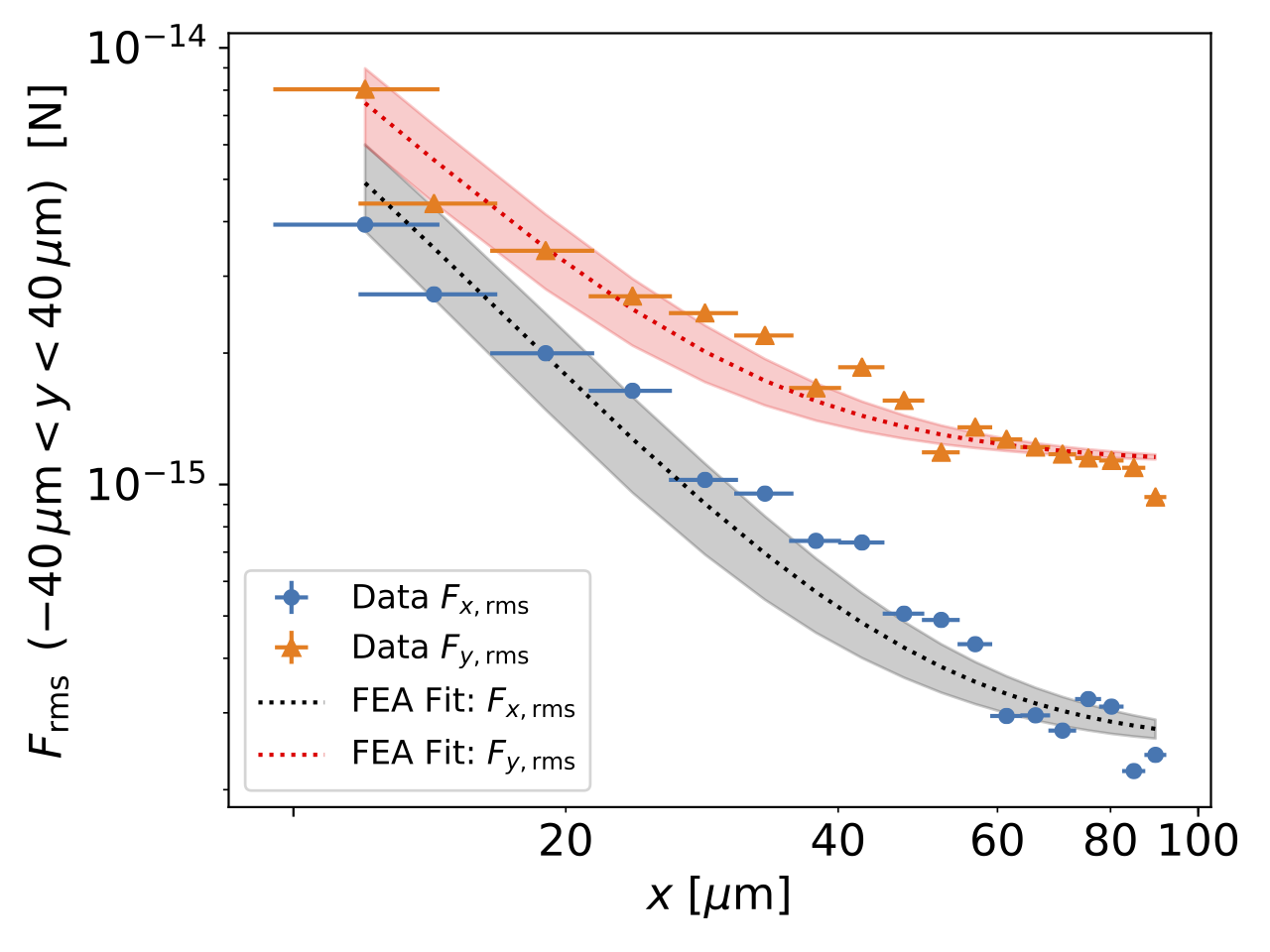}
\caption{ \footnotesize The rms of the $x$ and $y$ components of the measured force on the MS in the plane of the ACSB, as a function of separation from the ACSB. The data are compared to a model described in the text. The bands represents the standard deviation of the model, using different random implementations of the patch potentials.  The fit of $F_{x, {\rm rms}}$ ($F_{y, \rm{rms}}$) implies voltage patches of size ${\sim}0.8~\mu$m (${\sim}0.7~\mu$m), assuming $V_{\rm patch} = 100~$mV~\cite{SpeakeTrenkel:2003,Garret:2015}.}
\label{fig:patch_potentials_data}
\end{figure}

Patch potentials on the surface of the 200-nm-thick evaporated Au surface of the ACSB create an electric field which can be measured as a force on a charged MS. To perform this measurement, the ACSB is mechanically driven sinusoidally along the $y$ axis with the translation stage, over a regular grid of $x$ positions (separations), and $z$ positions. At each point within the $xz$ grid, $\vec{F}(t)$ and $y(t)$ are measured, so that the three-dimensional force field $\vec{F}(x,y,z)$ can be obtained. The relative registration of the MS with respect to the ACSB in terms of the three translations and six rotations is taken from the fit to the case of the biased ACSB, discussed in the previous section.  The three-dimensional electric field can then be extracted as $\vec{E}(x,y,z) = \vec{F}(x,y,z)/q$, where $q$ is the charge of the MS, also determined in the previous section.

The electric field due to patch potentials on the ACSB is numerically modeled following results obtained by Kelvin probe atomic force microscopy~\cite{Robertson:2006,Garret:2015}. Those authors define a voltage autocorrelation function, $R(\vec{s}) = \int \int d^{2}\vec{s}'V(\vec{s})V(\vec{s}' + \vec{s})$, where $\vec{s}$ and $\vec{s}'$ are positions on the surface. They then find that $R(\vec{s})$ is approximately constant for short length scales, followed by a sharp knee at a length scale consistent with the expected patch size, $l_{\rm patch}$. As an approximation to this autocorrelation function, we model triangular patches measuring $l_{\rm patch}$ on a side, with voltages randomly sampled from a normal distribution, $N(0,V_{\rm patch})$. FEA is used to determine the electric field due to these patches (as the geometry and boundary conditions do not permit an analytic solution), which we then compare to our data. Since the patches and resulting electric fields are random, we create many realizations of patches to sample the mean and variance of the root-mean-square (rms) force, $F_{\rm rms}$.

Measurements of $F_{\rm rms}$ along $x$ and $y$ for different separations $x$ are shown in Fig.~\ref{fig:patch_potentials_data}. The rms is computed over all grid points along the $y$ axis, at $z=0$.  In the figure, the data are fit to the results of the FEA model described above using a least-squared cost function, with a constant added variance to account for noise. Misalignments between the physical axes of the ACSB and the axes of the driven motion, together with an overall contact potential on the ACSB, would produce both a constant offset in $\vec{F}(t)$, as well as a term at the ACSB's driving frequency. Patch potentials with length scale $l_{\rm patch} \leq 40~\mu$m would only produce higher harmonics of the ACSB's driving frequency, as its maximum displacement is $(\Delta y)_{\rm max} = 80~\mu$m. Thus, when computing $\vec{F}(y)$ from $\vec{F}(t)$ and $y(t)$ via Fourier techniques, the dc and fundamental terms of the ACSB's driving frequency were set to zero in the FFT of $\vec{F}(t)$. At distances much greater than the patch size, the rms $\vec{E}$ field is proportional to the product of the patch length scale, $l_{\rm patch}$, and the rms patch voltage, $V_{\rm patch}$. Although the MS is not close enough to the ACSB to resolve the underlying patches, we are able to extract the patch length-scale voltage product. 

To compare to other measurements, we assume $V_{\rm patch}= 100~$mV~\cite{SpeakeTrenkel:2003,Garret:2015} and obtain $l_{\rm patch}$. Our fit of the $x$ rms ($y$ rms) force implies $l_{\rm patch} = 0.8~\mu$m ($l_{\rm patch} = 0.7~\mu$m). The length scales inferred from the fit are somewhat larger than the ${\sim}$200 nm grain size expected for a 200-nm-thick evaporated Au film~\cite{SpeakeTrenkel:2003,Garret:2015}. Others have observed sample contamination by dielectric adsorbates (such as silica dust) that become embedded in metallic surfaces, which can collect charge and affect measurements of patch potentials~\cite{Rossi:1992,Robertson:2006}. Finite electrode geometry may also play a role~\cite{Low:2011}.

\section{CONCLUSION}

We have demonstrated a technique capable of mapping, with sensitivity competitive to that of cryogenic AFM, a three-dimensional vector field over a volume of ${\sim}10^6$~$\mu$m$^3$ in free space.  The center of this sensor, a $4.7{\text -}\mu$m-diameter optically levitated dielectric microsphere, can be brought as close as $10~\mu$m to a metallic surface, while performing the measurement. This instrument is used to map the patch potentials on an Au surface.

While this result is exclusively sensitive to electric fields, three-dimensional mapping of other physical fields appears possible.  Magnetized microspheres are commercially available, and possibly neutral microspheres or microrods with large electric dipole moments could be used with enhanced sensitivity to electric field gradients.  Our group is actively pursuing the technique to search for new long-range interaction coupling to mass or other intrinsic properties of matter at the micrometer scale.

\section{ACKNOWLEDGEMENTS}

We would like to thank J.~Fox (Stanford) for discussions on the readout electronics, F.~Monteiro and D.~Moore (Yale) for general discussions related to trapping microspheres, and R.~DeVoe (Stanford) for a careful reading of the manuscript. This work was supported, in part, by NSF Grants No.~PHY1502156 and No.~PHY1802952 and the Heising-Simons Foundation.  A.K. acknowledges the partial support of the William~M. and Jane~D. Fairbank Postdoctoral Fellowship of Stanford University. Microsphere characterization was performed at the Stanford Nano Shared Facilities (SNSF), while ACSB fabrication was performed in part in the nano@Stanford labs, both of which are supported by the National Science Foundation as part of the National Nanotechnology Coordinated Infrastructure under Award No.~ECCS-1542152.

\bibliographystyle{apsrev4-1}
\bibliography{3d_force_microscopy}

%merlin.mbs apsrev4-1.bst 2010-07-25 4.21a (PWD, AO, DPC) hacked
%Control: key (0)
%Control: author (72) initials jnrlst
%Control: editor formatted (1) identically to author
%Control: production of article title (-1) disabled
%Control: page (0) single
%Control: year (1) truncated
%Control: production of eprint (0) enabled
\begin{thebibliography}{39}%
\makeatletter
\providecommand \@ifxundefined [1]{%
 \@ifx{#1\undefined}
}%
\providecommand \@ifnum [1]{%
 \ifnum #1\expandafter \@firstoftwo
 \else \expandafter \@secondoftwo
 \fi
}%
\providecommand \@ifx [1]{%
 \ifx #1\expandafter \@firstoftwo
 \else \expandafter \@secondoftwo
 \fi
}%
\providecommand \natexlab [1]{#1}%
\providecommand \enquote  [1]{``#1''}%
\providecommand \bibnamefont  [1]{#1}%
\providecommand \bibfnamefont [1]{#1}%
\providecommand \citenamefont [1]{#1}%
\providecommand \href@noop [0]{\@secondoftwo}%
\providecommand \href [0]{\begingroup \@sanitize@url \@href}%
\providecommand \@href[1]{\@@startlink{#1}\@@href}%
\providecommand \@@href[1]{\endgroup#1\@@endlink}%
\providecommand \@sanitize@url [0]{\catcode `\\12\catcode `\$12\catcode
  `\&12\catcode `\#12\catcode `\^12\catcode `\_12\catcode `\%12\relax}%
\providecommand \@@startlink[1]{}%
\providecommand \@@endlink[0]{}%
\providecommand \url  [0]{\begingroup\@sanitize@url \@url }%
\providecommand \@url [1]{\endgroup\@href {#1}{\urlprefix }}%
\providecommand \urlprefix  [0]{URL }%
\providecommand \Eprint [0]{\href }%
\providecommand \doibase [0]{http://dx.doi.org/}%
\providecommand \selectlanguage [0]{\@gobble}%
\providecommand \bibinfo  [0]{\@secondoftwo}%
\providecommand \bibfield  [0]{\@secondoftwo}%
\providecommand \translation [1]{[#1]}%
\providecommand \BibitemOpen [0]{}%
\providecommand \bibitemStop [0]{}%
\providecommand \bibitemNoStop [0]{.\EOS\space}%
\providecommand \EOS [0]{\spacefactor3000\relax}%
\providecommand \BibitemShut  [1]{\csname bibitem#1\endcsname}%
\let\auto@bib@innerbib\@empty
%</preamble>
\bibitem [{\citenamefont {Binnig}\ \emph {et~al.}(1986)\citenamefont {Binnig},
  \citenamefont {Quate},\ and\ \citenamefont {Gerber}}]{AFM:1986}%
  \BibitemOpen
  \bibfield  {author} {\bibinfo {author} {\bibfnamefont {G.}~\bibnamefont
  {Binnig}}, \bibinfo {author} {\bibfnamefont {C.~F.}\ \bibnamefont {Quate}}, \
  and\ \bibinfo {author} {\bibfnamefont {C.}~\bibnamefont {Gerber}},\ }\href
  {\doibase 10.1103/PhysRevLett.56.930} {\bibfield  {journal} {\bibinfo
  {journal} {Phys. Rev. Lett.}\ }\textbf {\bibinfo {volume} {56}},\ \bibinfo
  {pages} {930} (\bibinfo {year} {1986})}\BibitemShut {NoStop}%
\bibitem [{\citenamefont {Ashkin}(1970)}]{Ashkin:1970}%
  \BibitemOpen
  \bibfield  {author} {\bibinfo {author} {\bibfnamefont {A.}~\bibnamefont
  {Ashkin}},\ }\href {\doibase 10.1103/PhysRevLett.24.156} {\bibfield
  {journal} {\bibinfo  {journal} {Phys. Rev. Lett.}\ }\textbf {\bibinfo
  {volume} {24}},\ \bibinfo {pages} {156} (\bibinfo {year} {1970})}\BibitemShut
  {NoStop}%
\bibitem [{\citenamefont {{Ashkin}}\ and\ \citenamefont
  {{Dziedzic}}(1971)}]{Ashkin:1971}%
  \BibitemOpen
  \bibfield  {author} {\bibinfo {author} {\bibfnamefont {A.}~\bibnamefont
  {{Ashkin}}}\ and\ \bibinfo {author} {\bibfnamefont {J.~M.}\ \bibnamefont
  {{Dziedzic}}},\ }\href {\doibase 10.1063/1.1653919} {\bibfield  {journal}
  {\bibinfo  {journal} {Appl. Phys. Lett.}\ }\textbf {\bibinfo {volume} {19}},\
  \bibinfo {pages} {283} (\bibinfo {year} {1971})}\BibitemShut {NoStop}%
\bibitem [{\citenamefont {{Ashkin}}\ and\ \citenamefont
  {{Dziedzic}}(1977)}]{Ashkin:1977}%
  \BibitemOpen
  \bibfield  {author} {\bibinfo {author} {\bibfnamefont {A.}~\bibnamefont
  {{Ashkin}}}\ and\ \bibinfo {author} {\bibfnamefont {J.~M.}\ \bibnamefont
  {{Dziedzic}}},\ }\href {\doibase 10.1063/1.89335} {\bibfield  {journal}
  {\bibinfo  {journal} {Appl. Phys. Lett.}\ }\textbf {\bibinfo {volume} {30}},\
  \bibinfo {eid} {202} (\bibinfo {year} {1977})}\BibitemShut {NoStop}%
\bibitem [{\citenamefont {Bishop}\ \emph {et~al.}(2004)\citenamefont {Bishop},
  \citenamefont {Nieminen}, \citenamefont {Heckenberg},\ and\ \citenamefont
  {Rubinsztein-Dunlop}}]{Bishop:2004}%
  \BibitemOpen
  \bibfield  {author} {\bibinfo {author} {\bibfnamefont {A.~I.}\ \bibnamefont
  {Bishop}}, \bibinfo {author} {\bibfnamefont {T.~A.}\ \bibnamefont
  {Nieminen}}, \bibinfo {author} {\bibfnamefont {N.~R.}\ \bibnamefont
  {Heckenberg}}, \ and\ \bibinfo {author} {\bibfnamefont {H.}~\bibnamefont
  {Rubinsztein-Dunlop}},\ }\href {\doibase 10.1103/PhysRevLett.92.198104}
  {\bibfield  {journal} {\bibinfo  {journal} {Phys. Rev. Lett.}\ }\textbf
  {\bibinfo {volume} {92}},\ \bibinfo {pages} {198104} (\bibinfo {year}
  {2004})}\BibitemShut {NoStop}%
\bibitem [{\citenamefont {{Chang}}\ \emph {et~al.}(2010)\citenamefont
  {{Chang}}, \citenamefont {{Regal}}, \citenamefont {{Papp}}, \citenamefont
  {{Wilson}}, \citenamefont {{Ye}}, \citenamefont {{Painter}}, \citenamefont
  {{Kimble}},\ and\ \citenamefont {{Zoller}}}]{Chang:2010}%
  \BibitemOpen
  \bibfield  {author} {\bibinfo {author} {\bibfnamefont {D.~E.}\ \bibnamefont
  {{Chang}}}, \bibinfo {author} {\bibfnamefont {C.~A.}\ \bibnamefont
  {{Regal}}}, \bibinfo {author} {\bibfnamefont {S.~B.}\ \bibnamefont {{Papp}}},
  \bibinfo {author} {\bibfnamefont {D.~J.}\ \bibnamefont {{Wilson}}}, \bibinfo
  {author} {\bibfnamefont {J.}~\bibnamefont {{Ye}}}, \bibinfo {author}
  {\bibfnamefont {O.}~\bibnamefont {{Painter}}}, \bibinfo {author}
  {\bibfnamefont {H.~J.}\ \bibnamefont {{Kimble}}}, \ and\ \bibinfo {author}
  {\bibfnamefont {P.}~\bibnamefont {{Zoller}}},\ }\href {\doibase
  10.1073/pnas.0912969107} {\bibfield  {journal} {\bibinfo  {journal} {Proc.
  Nat. Acad. Sci. USA}\ }\textbf {\bibinfo {volume} {107}},\ \bibinfo {pages}
  {1005} (\bibinfo {year} {2010})}\BibitemShut {NoStop}%
\bibitem [{\citenamefont {Li}\ \emph {et~al.}(2011)\citenamefont {Li},
  \citenamefont {Kheifets},\ and\ \citenamefont {Raizen}}]{Li:2011}%
  \BibitemOpen
  \bibfield  {author} {\bibinfo {author} {\bibfnamefont {T.}~\bibnamefont
  {Li}}, \bibinfo {author} {\bibfnamefont {S.}~\bibnamefont {Kheifets}}, \ and\
  \bibinfo {author} {\bibfnamefont {M.~G.}\ \bibnamefont {Raizen}},\ }\href
  {\doibase 10.1038/nphys1952} {\bibfield  {journal} {\bibinfo  {journal} {Nat.
  Phys.}\ }\textbf {\bibinfo {volume} {7}},\ \bibinfo {pages} {527} (\bibinfo
  {year} {2011})}\BibitemShut {NoStop}%
%%CITATION = ARXIV:1101.1283;%%
\bibitem [{\citenamefont {Gieseler}\ \emph {et~al.}(2012)\citenamefont
  {Gieseler}, \citenamefont {Deutsch}, \citenamefont {Quidant},\ and\
  \citenamefont {Novotny}}]{Gieseler:2012}%
  \BibitemOpen
  \bibfield  {author} {\bibinfo {author} {\bibfnamefont {J.}~\bibnamefont
  {Gieseler}}, \bibinfo {author} {\bibfnamefont {B.}~\bibnamefont {Deutsch}},
  \bibinfo {author} {\bibfnamefont {R.}~\bibnamefont {Quidant}}, \ and\
  \bibinfo {author} {\bibfnamefont {L.}~\bibnamefont {Novotny}},\ }\href
  {\doibase 10.1103/PhysRevLett.109.103603} {\bibfield  {journal} {\bibinfo
  {journal} {Phys. Rev. Lett.}\ }\textbf {\bibinfo {volume} {109}},\ \bibinfo
  {pages} {103603} (\bibinfo {year} {2012})}\BibitemShut {NoStop}%
\bibitem [{\citenamefont {{Li}}()}]{Li:2013}%
  \BibitemOpen
  \bibfield  {author} {\bibinfo {author} {\bibfnamefont {T.}~\bibnamefont
  {{Li}}},\ }\href {\doibase 10.1007/978-1-4614-6031-2} {}\bibinfo {note}
  {Ph.D. {T}hesis, University of Texas at Austin, Austin, Texas, 2013
  (unpublished)}\BibitemShut {NoStop}%
\bibitem [{\citenamefont {Yin}\ \emph {et~al.}(2013)\citenamefont {Yin},
  \citenamefont {Geraci},\ and\ \citenamefont {Li}}]{Yin:2013}%
  \BibitemOpen
  \bibfield  {author} {\bibinfo {author} {\bibfnamefont {Z.-Q.}\ \bibnamefont
  {Yin}}, \bibinfo {author} {\bibfnamefont {A.~A.}\ \bibnamefont {Geraci}}, \
  and\ \bibinfo {author} {\bibfnamefont {T.}~\bibnamefont {Li}},\ }\href
  {\doibase 10.1142/S0217979213300181} {\bibfield  {journal} {\bibinfo
  {journal} {Int. J. Mod. Phys.}\ }\textbf {\bibinfo {volume} {B27}},\ \bibinfo
  {pages} {1330018} (\bibinfo {year} {2013})}\BibitemShut {NoStop}%
%%CITATION = ARXIV:1308.4503;%%
\bibitem [{\citenamefont {{Asenbaum}}\ \emph {et~al.}(2013)\citenamefont
  {{Asenbaum}}, \citenamefont {{Kuhn}}, \citenamefont {{Nimmrichter}},
  \citenamefont {{Sezer}},\ and\ \citenamefont {{Arndt}}}]{Asenbaum:2013}%
  \BibitemOpen
  \bibfield  {author} {\bibinfo {author} {\bibfnamefont {P.}~\bibnamefont
  {{Asenbaum}}}, \bibinfo {author} {\bibfnamefont {S.}~\bibnamefont {{Kuhn}}},
  \bibinfo {author} {\bibfnamefont {S.}~\bibnamefont {{Nimmrichter}}}, \bibinfo
  {author} {\bibfnamefont {U.}~\bibnamefont {{Sezer}}}, \ and\ \bibinfo
  {author} {\bibfnamefont {M.}~\bibnamefont {{Arndt}}},\ }\href {\doibase
  10.1038/ncomms3743} {\bibfield  {journal} {\bibinfo  {journal} {Nat.
  Commun.}\ }\textbf {\bibinfo {volume} {4}},\ \bibinfo {eid} {2743} (\bibinfo
  {year} {2013})}\BibitemShut {NoStop}%
\bibitem [{\citenamefont {Moore}\ \emph {et~al.}(2014)\citenamefont {Moore},
  \citenamefont {Rider},\ and\ \citenamefont {Gratta}}]{Moore:2014}%
  \BibitemOpen
  \bibfield  {author} {\bibinfo {author} {\bibfnamefont {D.~C.}\ \bibnamefont
  {Moore}}, \bibinfo {author} {\bibfnamefont {A.~D.}\ \bibnamefont {Rider}}, \
  and\ \bibinfo {author} {\bibfnamefont {G.}~\bibnamefont {Gratta}},\ }\href
  {\doibase 10.1103/PhysRevLett.113.251801} {\bibfield  {journal} {\bibinfo
  {journal} {Phys. Rev. Lett.}\ }\textbf {\bibinfo {volume} {113}},\ \bibinfo
  {pages} {251801} (\bibinfo {year} {2014})}\BibitemShut {NoStop}%
\bibitem [{\citenamefont {Ranjit}\ \emph {et~al.}(2015)\citenamefont {Ranjit},
  \citenamefont {Atherton}, \citenamefont {Stutz}, \citenamefont {Cunningham},\
  and\ \citenamefont {Geraci}}]{Ranjit:2015}%
  \BibitemOpen
  \bibfield  {author} {\bibinfo {author} {\bibfnamefont {G.}~\bibnamefont
  {Ranjit}}, \bibinfo {author} {\bibfnamefont {D.~P.}\ \bibnamefont
  {Atherton}}, \bibinfo {author} {\bibfnamefont {J.~H.}\ \bibnamefont {Stutz}},
  \bibinfo {author} {\bibfnamefont {M.}~\bibnamefont {Cunningham}}, \ and\
  \bibinfo {author} {\bibfnamefont {A.~A.}\ \bibnamefont {Geraci}},\ }\href
  {\doibase 10.1103/PhysRevA.91.051805} {\bibfield  {journal} {\bibinfo
  {journal} {Phys. Rev. A}\ }\textbf {\bibinfo {volume} {91}},\ \bibinfo
  {pages} {051805} (\bibinfo {year} {2015})}\BibitemShut {NoStop}%
\bibitem [{\citenamefont {Millen}\ \emph {et~al.}(2015)\citenamefont {Millen},
  \citenamefont {Fonseca}, \citenamefont {Mavrogordatos}, \citenamefont
  {Monteiro},\ and\ \citenamefont {Barker}}]{Millen:2015}%
  \BibitemOpen
  \bibfield  {author} {\bibinfo {author} {\bibfnamefont {J.}~\bibnamefont
  {Millen}}, \bibinfo {author} {\bibfnamefont {P.~Z.~G.}\ \bibnamefont
  {Fonseca}}, \bibinfo {author} {\bibfnamefont {T.}~\bibnamefont
  {Mavrogordatos}}, \bibinfo {author} {\bibfnamefont {T.~S.}\ \bibnamefont
  {Monteiro}}, \ and\ \bibinfo {author} {\bibfnamefont {P.~F.}\ \bibnamefont
  {Barker}},\ }\href {\doibase 10.1103/PhysRevLett.114.123602} {\bibfield
  {journal} {\bibinfo  {journal} {Phys. Rev. Lett.}\ }\textbf {\bibinfo
  {volume} {114}},\ \bibinfo {pages} {123602} (\bibinfo {year}
  {2015})}\BibitemShut {NoStop}%
\bibitem [{\citenamefont {Rider}\ \emph {et~al.}(2016)\citenamefont {Rider},
  \citenamefont {Moore}, \citenamefont {Blakemore}, \citenamefont {Louis},
  \citenamefont {Lu},\ and\ \citenamefont {Gratta}}]{Rider:2016}%
  \BibitemOpen
  \bibfield  {author} {\bibinfo {author} {\bibfnamefont {A.~D.}\ \bibnamefont
  {Rider}}, \bibinfo {author} {\bibfnamefont {D.~C.}\ \bibnamefont {Moore}},
  \bibinfo {author} {\bibfnamefont {C.~P.}\ \bibnamefont {Blakemore}}, \bibinfo
  {author} {\bibfnamefont {M.}~\bibnamefont {Louis}}, \bibinfo {author}
  {\bibfnamefont {M.}~\bibnamefont {Lu}}, \ and\ \bibinfo {author}
  {\bibfnamefont {G.}~\bibnamefont {Gratta}},\ }\href {\doibase
  10.1103/PhysRevLett.117.101101} {\bibfield  {journal} {\bibinfo  {journal}
  {Phys. Rev. Lett.}\ }\textbf {\bibinfo {volume} {117}},\ \bibinfo {pages}
  {101101} (\bibinfo {year} {2016})}\BibitemShut {NoStop}%
%%CITATION = ARXIV:1604.04908;%%
\bibitem [{\citenamefont {Jain}\ \emph {et~al.}(2016)\citenamefont {Jain},
  \citenamefont {Gieseler}, \citenamefont {Moritz}, \citenamefont {Dellago},
  \citenamefont {Quidant},\ and\ \citenamefont {Novotny}}]{Jain:2016}%
  \BibitemOpen
  \bibfield  {author} {\bibinfo {author} {\bibfnamefont {V.}~\bibnamefont
  {Jain}}, \bibinfo {author} {\bibfnamefont {J.}~\bibnamefont {Gieseler}},
  \bibinfo {author} {\bibfnamefont {C.}~\bibnamefont {Moritz}}, \bibinfo
  {author} {\bibfnamefont {C.}~\bibnamefont {Dellago}}, \bibinfo {author}
  {\bibfnamefont {R.}~\bibnamefont {Quidant}}, \ and\ \bibinfo {author}
  {\bibfnamefont {L.}~\bibnamefont {Novotny}},\ }\href {\doibase
  10.1103/PhysRevLett.116.243601} {\bibfield  {journal} {\bibinfo  {journal}
  {Phys. Rev. Lett.}\ }\textbf {\bibinfo {volume} {116}},\ \bibinfo {pages}
  {243601} (\bibinfo {year} {2016})}\BibitemShut {NoStop}%
\bibitem [{\citenamefont {Hoang}\ \emph {et~al.}(2016)\citenamefont {Hoang},
  \citenamefont {Ma}, \citenamefont {Ahn}, \citenamefont {Bang}, \citenamefont
  {Robicheaux}, \citenamefont {Yin},\ and\ \citenamefont {Li}}]{Hoang:2016}%
  \BibitemOpen
  \bibfield  {author} {\bibinfo {author} {\bibfnamefont {T.~M.}\ \bibnamefont
  {Hoang}}, \bibinfo {author} {\bibfnamefont {Y.}~\bibnamefont {Ma}}, \bibinfo
  {author} {\bibfnamefont {J.}~\bibnamefont {Ahn}}, \bibinfo {author}
  {\bibfnamefont {J.}~\bibnamefont {Bang}}, \bibinfo {author} {\bibfnamefont
  {F.}~\bibnamefont {Robicheaux}}, \bibinfo {author} {\bibfnamefont {Z.-Q.}\
  \bibnamefont {Yin}}, \ and\ \bibinfo {author} {\bibfnamefont
  {T.}~\bibnamefont {Li}},\ }\href {\doibase 10.1103/PhysRevLett.117.123604}
  {\bibfield  {journal} {\bibinfo  {journal} {Phys. Rev. Lett.}\ }\textbf
  {\bibinfo {volume} {117}},\ \bibinfo {pages} {123604} (\bibinfo {year}
  {2016})}\BibitemShut {NoStop}%
\bibitem [{\citenamefont {Ranjit}\ \emph {et~al.}(2016)\citenamefont {Ranjit},
  \citenamefont {Cunningham}, \citenamefont {Casey},\ and\ \citenamefont
  {Geraci}}]{Ranjit:2016}%
  \BibitemOpen
  \bibfield  {author} {\bibinfo {author} {\bibfnamefont {G.}~\bibnamefont
  {Ranjit}}, \bibinfo {author} {\bibfnamefont {M.}~\bibnamefont {Cunningham}},
  \bibinfo {author} {\bibfnamefont {K.}~\bibnamefont {Casey}}, \ and\ \bibinfo
  {author} {\bibfnamefont {A.~A.}\ \bibnamefont {Geraci}},\ }\href {\doibase
  10.1103/PhysRevA.93.053801} {\bibfield  {journal} {\bibinfo  {journal} {Phys.
  Rev. A}\ }\textbf {\bibinfo {volume} {93}},\ \bibinfo {pages} {053801}
  (\bibinfo {year} {2016})}\BibitemShut {NoStop}%
\bibitem [{\citenamefont {Mazilu}\ \emph {et~al.}(2016)\citenamefont {Mazilu},
  \citenamefont {Arita}, \citenamefont {Vettenburg}, \citenamefont {Au\~n\'on},
  \citenamefont {Wright},\ and\ \citenamefont {Dholakia}}]{Mazilu:2016}%
  \BibitemOpen
  \bibfield  {author} {\bibinfo {author} {\bibfnamefont {M.}~\bibnamefont
  {Mazilu}}, \bibinfo {author} {\bibfnamefont {Y.}~\bibnamefont {Arita}},
  \bibinfo {author} {\bibfnamefont {T.}~\bibnamefont {Vettenburg}}, \bibinfo
  {author} {\bibfnamefont {J.~M.}\ \bibnamefont {Au\~n\'on}}, \bibinfo {author}
  {\bibfnamefont {E.~M.}\ \bibnamefont {Wright}}, \ and\ \bibinfo {author}
  {\bibfnamefont {K.}~\bibnamefont {Dholakia}},\ }\href {\doibase
  10.1103/PhysRevA.94.053821} {\bibfield  {journal} {\bibinfo  {journal} {Phys.
  Rev. A}\ }\textbf {\bibinfo {volume} {94}},\ \bibinfo {pages} {053821}
  (\bibinfo {year} {2016})}\BibitemShut {NoStop}%
\bibitem [{\citenamefont {Fonseca}\ \emph {et~al.}(2016)\citenamefont
  {Fonseca}, \citenamefont {Aranas}, \citenamefont {Millen}, \citenamefont
  {Monteiro},\ and\ \citenamefont {Barker}}]{Fonseca:2016}%
  \BibitemOpen
  \bibfield  {author} {\bibinfo {author} {\bibfnamefont {P.~Z.~G.}\
  \bibnamefont {Fonseca}}, \bibinfo {author} {\bibfnamefont {E.~B.}\
  \bibnamefont {Aranas}}, \bibinfo {author} {\bibfnamefont {J.}~\bibnamefont
  {Millen}}, \bibinfo {author} {\bibfnamefont {T.~S.}\ \bibnamefont
  {Monteiro}}, \ and\ \bibinfo {author} {\bibfnamefont {P.~F.}\ \bibnamefont
  {Barker}},\ }\href {\doibase 10.1103/PhysRevLett.117.173602} {\bibfield
  {journal} {\bibinfo  {journal} {Phys. Rev. Lett.}\ }\textbf {\bibinfo
  {volume} {117}},\ \bibinfo {pages} {173602} (\bibinfo {year}
  {2016})}\BibitemShut {NoStop}%
\bibitem [{\citenamefont {{Vovrosh}}\ \emph {et~al.}(2017)\citenamefont
  {{Vovrosh}}, \citenamefont {{Rashid}}, \citenamefont {{Hempston}},
  \citenamefont {{Bateman}}, \citenamefont {{Paternostro}},\ and\ \citenamefont
  {{Ulbricht}}}]{Vovrosh:2017}%
  \BibitemOpen
  \bibfield  {author} {\bibinfo {author} {\bibfnamefont {J.}~\bibnamefont
  {{Vovrosh}}}, \bibinfo {author} {\bibfnamefont {M.}~\bibnamefont {{Rashid}}},
  \bibinfo {author} {\bibfnamefont {D.}~\bibnamefont {{Hempston}}}, \bibinfo
  {author} {\bibfnamefont {J.}~\bibnamefont {{Bateman}}}, \bibinfo {author}
  {\bibfnamefont {M.}~\bibnamefont {{Paternostro}}}, \ and\ \bibinfo {author}
  {\bibfnamefont {H.}~\bibnamefont {{Ulbricht}}},\ }\href {\doibase
  10.1364/JOSAB.34.001421} {\bibfield  {journal} {\bibinfo  {journal} {J. Opt.
  Soc. Am. B}\ }\textbf {\bibinfo {volume} {34}},\ \bibinfo {pages} {1421}
  (\bibinfo {year} {2017})}\BibitemShut {NoStop}%
\bibitem [{\citenamefont {Monteiro}\ \emph {et~al.}(2017)\citenamefont
  {Monteiro}, \citenamefont {Ghosh}, \citenamefont {Fine},\ and\ \citenamefont
  {Moore}}]{Monteiro:2017}%
  \BibitemOpen
  \bibfield  {author} {\bibinfo {author} {\bibfnamefont {F.}~\bibnamefont
  {Monteiro}}, \bibinfo {author} {\bibfnamefont {S.}~\bibnamefont {Ghosh}},
  \bibinfo {author} {\bibfnamefont {A.~G.}\ \bibnamefont {Fine}}, \ and\
  \bibinfo {author} {\bibfnamefont {D.~C.}\ \bibnamefont {Moore}},\ }\href
  {\doibase 10.1103/PhysRevA.96.063841} {\bibfield  {journal} {\bibinfo
  {journal} {Phys. Rev. A}\ }\textbf {\bibinfo {volume} {96}},\ \bibinfo
  {pages} {063841} (\bibinfo {year} {2017})}\BibitemShut {NoStop}%
\bibitem [{\citenamefont {Hempston}\ \emph {et~al.}(2017)\citenamefont
  {Hempston}, \citenamefont {Vovrosh}, \citenamefont {Toroš}, \citenamefont
  {Winstone}, \citenamefont {Rashid},\ and\ \citenamefont
  {Ulbricht}}]{Hempston:2017}%
  \BibitemOpen
  \bibfield  {author} {\bibinfo {author} {\bibfnamefont {D.}~\bibnamefont
  {Hempston}}, \bibinfo {author} {\bibfnamefont {J.}~\bibnamefont {Vovrosh}},
  \bibinfo {author} {\bibfnamefont {M.}~\bibnamefont {Toroš}}, \bibinfo
  {author} {\bibfnamefont {G.}~\bibnamefont {Winstone}}, \bibinfo {author}
  {\bibfnamefont {M.}~\bibnamefont {Rashid}}, \ and\ \bibinfo {author}
  {\bibfnamefont {H.}~\bibnamefont {Ulbricht}},\ }\href {\doibase
  10.1063/1.4993555} {\bibfield  {journal} {\bibinfo  {journal} {Appl. Phys.
  Lett.}\ }\textbf {\bibinfo {volume} {111}},\ \bibinfo {pages} {133111}
  (\bibinfo {year} {2017})}\BibitemShut {NoStop}%
\bibitem [{\citenamefont {Rider}\ \emph {et~al.}(2018)\citenamefont {Rider},
  \citenamefont {Blakemore}, \citenamefont {Gratta},\ and\ \citenamefont
  {Moore}}]{Rider:2018}%
  \BibitemOpen
  \bibfield  {author} {\bibinfo {author} {\bibfnamefont {A.~D.}\ \bibnamefont
  {Rider}}, \bibinfo {author} {\bibfnamefont {C.~P.}\ \bibnamefont
  {Blakemore}}, \bibinfo {author} {\bibfnamefont {G.}~\bibnamefont {Gratta}}, \
  and\ \bibinfo {author} {\bibfnamefont {D.~C.}\ \bibnamefont {Moore}},\ }\href
  {\doibase 10.1103/PhysRevA.97.013842} {\bibfield  {journal} {\bibinfo
  {journal} {Phys. Rev. A}\ }\textbf {\bibinfo {volume} {97}},\ \bibinfo
  {pages} {013842} (\bibinfo {year} {2018})}\BibitemShut {NoStop}%
\bibitem [{\citenamefont {Monteiro}\ \emph {et~al.}(2018)\citenamefont
  {Monteiro}, \citenamefont {Ghosh}, \citenamefont {van Assendelft},\ and\
  \citenamefont {Moore}}]{Monteiro:2018}%
  \BibitemOpen
  \bibfield  {author} {\bibinfo {author} {\bibfnamefont {F.}~\bibnamefont
  {Monteiro}}, \bibinfo {author} {\bibfnamefont {S.}~\bibnamefont {Ghosh}},
  \bibinfo {author} {\bibfnamefont {E.~C.}\ \bibnamefont {van Assendelft}}, \
  and\ \bibinfo {author} {\bibfnamefont {D.~C.}\ \bibnamefont {Moore}},\ }\href
  {\doibase 10.1103/PhysRevA.97.051802} {\bibfield  {journal} {\bibinfo
  {journal} {Phys. Rev. A}\ }\textbf {\bibinfo {volume} {97}},\ \bibinfo
  {pages} {051802} (\bibinfo {year} {2018})}\BibitemShut {NoStop}%
\bibitem [{\citenamefont {Diehl}\ \emph {et~al.}(2018)\citenamefont {Diehl},
  \citenamefont {Hebestreit}, \citenamefont {Reimann}, \citenamefont
  {Tebbenjohanns}, \citenamefont {Frimmer},\ and\ \citenamefont
  {Novotny}}]{Diehl:2018}%
  \BibitemOpen
  \bibfield  {author} {\bibinfo {author} {\bibfnamefont {R.}~\bibnamefont
  {Diehl}}, \bibinfo {author} {\bibfnamefont {E.}~\bibnamefont {Hebestreit}},
  \bibinfo {author} {\bibfnamefont {R.}~\bibnamefont {Reimann}}, \bibinfo
  {author} {\bibfnamefont {F.}~\bibnamefont {Tebbenjohanns}}, \bibinfo {author}
  {\bibfnamefont {M.}~\bibnamefont {Frimmer}}, \ and\ \bibinfo {author}
  {\bibfnamefont {L.}~\bibnamefont {Novotny}},\ }\href {\doibase
  10.1103/PhysRevA.98.013851} {\bibfield  {journal} {\bibinfo  {journal} {Phys.
  Rev. A}\ }\textbf {\bibinfo {volume} {98}},\ \bibinfo {pages} {013851}
  (\bibinfo {year} {2018})}\BibitemShut {NoStop}%
\bibitem [{\citenamefont {Geraci}\ \emph {et~al.}(2010)\citenamefont {Geraci},
  \citenamefont {Papp},\ and\ \citenamefont {Kitching}}]{Geraci:2010}%
  \BibitemOpen
  \bibfield  {author} {\bibinfo {author} {\bibfnamefont {A.~A.}\ \bibnamefont
  {Geraci}}, \bibinfo {author} {\bibfnamefont {S.~B.}\ \bibnamefont {Papp}}, \
  and\ \bibinfo {author} {\bibfnamefont {J.}~\bibnamefont {Kitching}},\ }\href
  {\doibase 10.1103/PhysRevLett.105.101101} {\bibfield  {journal} {\bibinfo
  {journal} {Phys. Rev. Lett.}\ }\textbf {\bibinfo {volume} {105}},\ \bibinfo
  {pages} {101101} (\bibinfo {year} {2010})}\BibitemShut {NoStop}%
\bibitem [{\citenamefont {Ether.}\ \emph {et~al.}(2015)\citenamefont {Ether.},
  \citenamefont {Pires}, \citenamefont {Umrath}, \citenamefont {Martinez},
  \citenamefont {Ayala}, \citenamefont {Pontes}, \citenamefont {de~S.~Araújo},
  \citenamefont {Frases}, \citenamefont {Ingold}, \citenamefont {Rosa},
  \citenamefont {Viana}, \citenamefont {Nussenzveig},\ and\ \citenamefont
  {Neto}}]{Ether:2015}%
  \BibitemOpen
  \bibfield  {author} {\bibinfo {author} {\bibfnamefont {D.~S.}\ \bibnamefont
  {Ether.}}, \bibinfo {author} {\bibfnamefont {L.~B.}\ \bibnamefont {Pires}},
  \bibinfo {author} {\bibfnamefont {S.}~\bibnamefont {Umrath}}, \bibinfo
  {author} {\bibfnamefont {D.}~\bibnamefont {Martinez}}, \bibinfo {author}
  {\bibfnamefont {Y.}~\bibnamefont {Ayala}}, \bibinfo {author} {\bibfnamefont
  {B.}~\bibnamefont {Pontes}}, \bibinfo {author} {\bibfnamefont {G.~R.}\
  \bibnamefont {de~S.~Araújo}}, \bibinfo {author} {\bibfnamefont
  {S.}~\bibnamefont {Frases}}, \bibinfo {author} {\bibfnamefont {G.-L.}\
  \bibnamefont {Ingold}}, \bibinfo {author} {\bibfnamefont {F.~S.~S.}\
  \bibnamefont {Rosa}}, \bibinfo {author} {\bibfnamefont {N.~B.}\ \bibnamefont
  {Viana}}, \bibinfo {author} {\bibfnamefont {H.~M.}\ \bibnamefont
  {Nussenzveig}}, \ and\ \bibinfo {author} {\bibfnamefont {P.~A.~M.}\
  \bibnamefont {Neto}},\ }\href
  {http://stacks.iop.org/0295-5075/112/i=4/a=44001} {\bibfield  {journal}
  {\bibinfo  {journal} {EPL}\ }\textbf {\bibinfo {volume} {112}},\ \bibinfo
  {pages} {44001} (\bibinfo {year} {2015})}\BibitemShut {NoStop}%
\bibitem [{\citenamefont {Winstone}\ \emph {et~al.}(2018)\citenamefont
  {Winstone}, \citenamefont {Bennett}, \citenamefont {Rademacher},
  \citenamefont {Rashid}, \citenamefont {Buhmann},\ and\ \citenamefont
  {Ulbricht}}]{Winstone:2018}%
  \BibitemOpen
  \bibfield  {author} {\bibinfo {author} {\bibfnamefont {G.}~\bibnamefont
  {Winstone}}, \bibinfo {author} {\bibfnamefont {R.}~\bibnamefont {Bennett}},
  \bibinfo {author} {\bibfnamefont {M.}~\bibnamefont {Rademacher}}, \bibinfo
  {author} {\bibfnamefont {M.}~\bibnamefont {Rashid}}, \bibinfo {author}
  {\bibfnamefont {S.}~\bibnamefont {Buhmann}}, \ and\ \bibinfo {author}
  {\bibfnamefont {H.}~\bibnamefont {Ulbricht}},\ }\href {\doibase
  10.1103/PhysRevA.98.053831} {\bibfield  {journal} {\bibinfo  {journal} {Phys.
  Rev. A}\ }\textbf {\bibinfo {volume} {98}},\ \bibinfo {pages} {053831}
  (\bibinfo {year} {2018})}\BibitemShut {NoStop}%
\bibitem [{\citenamefont {\relax Bangs~Laboratories}()}]{bangs_laboratories}%
  \BibitemOpen
  \bibfield  {author} {\bibinfo {author} {\bibnamefont {\relax
  Bangs~Laboratories}},\ }\href@noop {} {}\bibinfo {howpublished}
  {\url{http://www.bangslabs.com/}},\ \bibinfo {note} {accessed:
  2016-04-16}\BibitemShut {NoStop}%
\bibitem [{\citenamefont {Wang}\ \emph {et~al.}(2017)\citenamefont {Wang},
  \citenamefont {Rider}, \citenamefont {Moore}, \citenamefont {Blakemore},
  \citenamefont {Cao},\ and\ \citenamefont {Gratta}}]{Wang:2017}%
  \BibitemOpen
  \bibfield  {author} {\bibinfo {author} {\bibfnamefont {Q.}~\bibnamefont
  {Wang}}, \bibinfo {author} {\bibfnamefont {A.~D.}\ \bibnamefont {Rider}},
  \bibinfo {author} {\bibfnamefont {D.~C.}\ \bibnamefont {Moore}}, \bibinfo
  {author} {\bibfnamefont {C.~P.}\ \bibnamefont {Blakemore}}, \bibinfo {author}
  {\bibfnamefont {L.}~\bibnamefont {Cao}}, \ and\ \bibinfo {author}
  {\bibfnamefont {G.}~\bibnamefont {Gratta}},\ }\href {\doibase
  10.1109/ECTC.2017.274} {\bibfield  {journal} {\bibinfo  {journal} {Proc. IEEE
  ECTC}\ ,\ \bibinfo {pages} {1773}} (\bibinfo {year} {2017})}\BibitemShut
  {NoStop}%
\bibitem [{\citenamefont {Turchette}\ \emph {et~al.}(2000)\citenamefont
  {Turchette}, \citenamefont {Kielpinski}, \citenamefont {King}, \citenamefont
  {Leibfried}, \citenamefont {Meekhof}, \citenamefont {Myatt}, \citenamefont
  {Rowe}, \citenamefont {Sackett}, \citenamefont {Wood}, \citenamefont {Itano},
  \citenamefont {Monroe},\ and\ \citenamefont {Wineland}}]{Turchette:2000}%
  \BibitemOpen
  \bibfield  {author} {\bibinfo {author} {\bibfnamefont {Q.~A.}\ \bibnamefont
  {Turchette}}, \bibinfo {author} {\bibfnamefont {D.}~\bibnamefont
  {Kielpinski}}, \bibinfo {author} {\bibfnamefont {B.~E.}\ \bibnamefont
  {King}}, \bibinfo {author} {\bibfnamefont {D.}~\bibnamefont {Leibfried}},
  \bibinfo {author} {\bibfnamefont {D.~M.}\ \bibnamefont {Meekhof}}, \bibinfo
  {author} {\bibfnamefont {C.~J.}\ \bibnamefont {Myatt}}, \bibinfo {author}
  {\bibfnamefont {M.~A.}\ \bibnamefont {Rowe}}, \bibinfo {author}
  {\bibfnamefont {C.~A.}\ \bibnamefont {Sackett}}, \bibinfo {author}
  {\bibfnamefont {C.~S.}\ \bibnamefont {Wood}}, \bibinfo {author}
  {\bibfnamefont {W.~M.}\ \bibnamefont {Itano}}, \bibinfo {author}
  {\bibfnamefont {C.}~\bibnamefont {Monroe}}, \ and\ \bibinfo {author}
  {\bibfnamefont {D.~J.}\ \bibnamefont {Wineland}},\ }\href {\doibase
  10.1103/PhysRevA.61.063418} {\bibfield  {journal} {\bibinfo  {journal} {Phys.
  Rev. A}\ }\textbf {\bibinfo {volume} {61}},\ \bibinfo {pages} {063418}
  (\bibinfo {year} {2000})}\BibitemShut {NoStop}%
\bibitem [{\citenamefont {Low}\ \emph {et~al.}(2011)\citenamefont {Low},
  \citenamefont {Herskind},\ and\ \citenamefont {Chuang}}]{Low:2011}%
  \BibitemOpen
  \bibfield  {author} {\bibinfo {author} {\bibfnamefont {G.~H.}\ \bibnamefont
  {Low}}, \bibinfo {author} {\bibfnamefont {P.~F.}\ \bibnamefont {Herskind}}, \
  and\ \bibinfo {author} {\bibfnamefont {I.~L.}\ \bibnamefont {Chuang}},\
  }\href {\doibase 10.1103/PhysRevA.84.053425} {\bibfield  {journal} {\bibinfo
  {journal} {Phys. Rev. A}\ }\textbf {\bibinfo {volume} {84}},\ \bibinfo
  {pages} {053425} (\bibinfo {year} {2011})}\BibitemShut {NoStop}%
\bibitem [{\citenamefont {Chiaverini}\ \emph {et~al.}(2003)\citenamefont
  {Chiaverini}, \citenamefont {Smullin}, \citenamefont {Geraci}, \citenamefont
  {Weld},\ and\ \citenamefont {Kapitulnik}}]{Chiaverini:2003}%
  \BibitemOpen
  \bibfield  {author} {\bibinfo {author} {\bibfnamefont {J.}~\bibnamefont
  {Chiaverini}}, \bibinfo {author} {\bibfnamefont {S.~J.}\ \bibnamefont
  {Smullin}}, \bibinfo {author} {\bibfnamefont {A.~A.}\ \bibnamefont {Geraci}},
  \bibinfo {author} {\bibfnamefont {D.~M.}\ \bibnamefont {Weld}}, \ and\
  \bibinfo {author} {\bibfnamefont {A.}~\bibnamefont {Kapitulnik}},\ }\href
  {\doibase 10.1103/PhysRevLett.90.151101} {\bibfield  {journal} {\bibinfo
  {journal} {Phys. Rev. Lett.}\ }\textbf {\bibinfo {volume} {90}},\ \bibinfo
  {pages} {151101} (\bibinfo {year} {2003})}\BibitemShut {NoStop}%
\bibitem [{\citenamefont {Churnside}\ and\ \citenamefont
  {Perkins}(2014)}]{Churnside:2014}%
  \BibitemOpen
  \bibfield  {author} {\bibinfo {author} {\bibfnamefont {A.~B.}\ \bibnamefont
  {Churnside}}\ and\ \bibinfo {author} {\bibfnamefont {T.~T.}\ \bibnamefont
  {Perkins}},\ }\href {\doibase 10.1016/j.febslet.2014.04.033} {\bibfield
  {journal} {\bibinfo  {journal} {FEBS Lett.}\ }\textbf {\bibinfo {volume}
  {588}},\ \bibinfo {pages} {3621} (\bibinfo {year} {2014})}\BibitemShut
  {NoStop}%
\bibitem [{\citenamefont {Rossi}\ and\ \citenamefont
  {Opat}(1992)}]{Rossi:1992}%
  \BibitemOpen
  \bibfield  {author} {\bibinfo {author} {\bibfnamefont {F.}~\bibnamefont
  {Rossi}}\ and\ \bibinfo {author} {\bibfnamefont {G.~I.}\ \bibnamefont
  {Opat}},\ }\href {http://stacks.iop.org/0022-3727/25/i=9/a=012} {\bibfield
  {journal} {\bibinfo  {journal} {Journal of Physics D: Applied Physics}\
  }\textbf {\bibinfo {volume} {25}},\ \bibinfo {pages} {1349} (\bibinfo {year}
  {1992})}\BibitemShut {NoStop}%
\bibitem [{\citenamefont {Robertson}\ \emph {et~al.}(2006)\citenamefont
  {Robertson}, \citenamefont {Blackwood}, \citenamefont {Buchman},
  \citenamefont {Byer}, \citenamefont {Camp}, \citenamefont {Gill},
  \citenamefont {Hanson}, \citenamefont {Williams},\ and\ \citenamefont
  {Zhou}}]{Robertson:2006}%
  \BibitemOpen
  \bibfield  {author} {\bibinfo {author} {\bibfnamefont {N.~A.}\ \bibnamefont
  {Robertson}}, \bibinfo {author} {\bibfnamefont {J.~R.}\ \bibnamefont
  {Blackwood}}, \bibinfo {author} {\bibfnamefont {S.}~\bibnamefont {Buchman}},
  \bibinfo {author} {\bibfnamefont {R.~L.}\ \bibnamefont {Byer}}, \bibinfo
  {author} {\bibfnamefont {J.}~\bibnamefont {Camp}}, \bibinfo {author}
  {\bibfnamefont {D.}~\bibnamefont {Gill}}, \bibinfo {author} {\bibfnamefont
  {J.}~\bibnamefont {Hanson}}, \bibinfo {author} {\bibfnamefont
  {S.}~\bibnamefont {Williams}}, \ and\ \bibinfo {author} {\bibfnamefont
  {P.}~\bibnamefont {Zhou}},\ }\href
  {http://stacks.iop.org/0264-9381/23/i=7/a=026} {\bibfield  {journal}
  {\bibinfo  {journal} {Classical and Quantum Gravity}\ }\textbf {\bibinfo
  {volume} {23}},\ \bibinfo {pages} {2665} (\bibinfo {year}
  {2006})}\BibitemShut {NoStop}%
\bibitem [{\citenamefont {Speake}\ and\ \citenamefont
  {Trenkel}(2003)}]{SpeakeTrenkel:2003}%
  \BibitemOpen
  \bibfield  {author} {\bibinfo {author} {\bibfnamefont {C.~C.}\ \bibnamefont
  {Speake}}\ and\ \bibinfo {author} {\bibfnamefont {C.}~\bibnamefont
  {Trenkel}},\ }\href {\doibase 10.1103/PhysRevLett.90.160403} {\bibfield
  {journal} {\bibinfo  {journal} {Phys. Rev. Lett.}\ }\textbf {\bibinfo
  {volume} {90}},\ \bibinfo {pages} {160403} (\bibinfo {year}
  {2003})}\BibitemShut {NoStop}%
\bibitem [{\citenamefont {Garrett}\ \emph {et~al.}(2015)\citenamefont
  {Garrett}, \citenamefont {Somers},\ and\ \citenamefont
  {Munday}}]{Garret:2015}%
  \BibitemOpen
  \bibfield  {author} {\bibinfo {author} {\bibfnamefont {J.~L.}\ \bibnamefont
  {Garrett}}, \bibinfo {author} {\bibfnamefont {D.}~\bibnamefont {Somers}}, \
  and\ \bibinfo {author} {\bibfnamefont {J.~N.}\ \bibnamefont {Munday}},\
  }\href {http://stacks.iop.org/0953-8984/27/i=21/a=214012} {\bibfield
  {journal} {\bibinfo  {journal} {Journal of Physics: Condensed Matter}\
  }\textbf {\bibinfo {volume} {27}},\ \bibinfo {pages} {214012} (\bibinfo
  {year} {2015})}\BibitemShut {NoStop}%
\end{thebibliography}%

\end{document}